\documentclass[aps,pra,reprint,groupedaddress,showpacs]{revtex4-1}

\usepackage{graphicx}
\usepackage{amsmath,amssymb,amsthm,latexsym}

\usepackage[utf8]{inputenc}

\usepackage{color} 
\usepackage{hyperref}

\newcommand{\etal}{\textit{et al.}}

\begin{document}

\normalsize
\title{Effect of the iron valence in the two types of layers in LiFeO$_2$Fe$_2$Se$_2$}

\author{Christoph Heil}
\email[]{christoph.heil@tugraz.at}
\affiliation{Institute of Theoretical and Computational Physics, University of 
Technology Graz, 8010 Graz, Austria}

\author{Lilia Boeri}
\affiliation{Institute of Theoretical and Computational Physics, University of 
Technology Graz, 8010 Graz, Austria}

\author{Heinrich Sormann}
\affiliation{Institute of Theoretical and Computational Physics, University of 
Technology Graz, 8010 Graz, Austria}

\author{Wolfgang von der Linden}
\affiliation{Institute of Theoretical and Computational Physics, University of 
Technology Graz, 8010 Graz, Austria}

\author{Markus Aichhorn}
\affiliation{Institute of Theoretical and Computational Physics, University of 
Technology Graz, 8010 Graz, Austria}

\date{\today}

\begin{abstract}
We perform electronic structure calculations for the recently
synthesized iron-based superconductor LiFeO$_2$Fe$_2$Se$_2$. 
In contrast to other iron-based
superconductors, this material comprises two different iron atoms in 3$d^5$ and 3$d^6$
configurations. In band theory, both contribute to the low-energy
electronic structure.
Spin-polarized density functional theory calculations predict an antiferromagnetic metallic ground
state with different moments on the two Fe sites. 
However, several other almost degenerate magnetic configurations exist.
Due to their different valences, the two iron atoms behave very differently when local
quantum correlations are included through the dynamical mean-field theory.
The contributions from the half-filled 3$d^5$ atoms in the LiFeO$_2$ layer are suppressed and the 3$d^6$ states from the FeSe layer restore the 
standard iron-based superconductor fermiology. 
\end{abstract}


\maketitle

\section{Introduction}
\label{sec:introduction}

The discovery of 
iron-based high-T$_c$ superconductors
(FeSCs) in 2008~\cite{kamihara_iron-based_2008} has
triggered an enormeous amount
of research in solid state physics, both
experimental and theoretical. Since then, many new compounds have
been discovered and investigated, which differ considerably in their
structural details~\cite{johnston_puzzle_2010}.
They all share a common structural motif, i.e., a square lattice of Fe atoms to which pnictogen or
chalcogen atoms are tetrahedrally coordinated. In the case of
pnictogen compounds, spacer layers such as layers of lanthanide
oxides~\cite{jesche_strong_2008,carlo_static_2009,drew_coexistence_2009}
or alkaline (earth)
atoms \cite{huang_neutron-diffraction_2008,chu_synthesis_2009,dagotto_colloquium:_2013} between the Fe planes ensure charge neutrality. 
In chalcogenide compounds, these layers are
not necessary, leading to the structurally simplest iron-based
superconductors FeSe and FeTe.

Recently, Lu~\etal~\cite{lu_superconductivity_2014} synthesized
LiFeO$_2$Fe$_2$Se$_2$, i.e., a FeSe compound, intercalated with
LiFeO$_2$ layers, and reported a very high critical temperature of
$T_c=43$\,K, comparable to that of
many other high-T$_c$ FeSCs, including (mol)-FeSe~\cite{kamihara_iron-based_2008,rotter_superconductivity_2008,margadonna_pressure_2009, burrard-lucas_enhancement_2013}.
Together with Sr$_2$VO$_3$FeAs~\cite{zhu_transition_2009}, and
rare-earth (Ce,Pr,Eu) 1111 pnictides, this is one of the few examples of
a FeSC with a magnetic atom in the intercalated layers.     
What makes this compound special is that 
the magnetic atom is itself iron.
The iron atoms in the LiFeO$_2$ and FeSe layers, Fe$_{\text{Li}}$
and Fe$_{\text{Se}}$, respectively, have very different
properties related to their nominal charge. Since Fe$_{\text{Li}}$ is in
a 3$d^5$ configuration and therefore at half filling, correlations are
very effective and may lead to a Mott-insulating state for
moderate values of the Coulomb interaction $U$ and Hund's coupling
$J$~\cite{de_medici_janus-faced_2011}.
Fe$_{\text{Se}}$, on the other hand, is in 3$d^6$ configuration, 
i.e., well into the  Hund's metal
regime~\cite{yin_kinetic_2011,haule_coherenceincoherence_2009,georges_strong_2013},
where the correlated metallic state extends to much larger values of
the Coulomb interaction $U$. 
This compound could thus provide two {\em qualitatively} different 
realizations of correlation effects due to Hund's coupling in one and
the same compound.

The electronic structure of  LiFeO$_2$Fe$_2$Se$_2$ was studied in 
Ref.~\onlinecite{liu_coexistence_2014} with standard density functional
calculations. The authors found an antiferromagnetic (AFM) ground state
in which both layers are metallic, and argued that 
with the inclusion of local correlations the LiFeO$_2$ plane
would become insulating, while the FeSe layer
would exhibit a bad metallic behavior.  
However, no quantitative evidence for this argument was provided.

In this paper, we study the electronic 
structure of LiFeO$_2$Fe$_2$Se$_2$ 
{\em including} strong local correlations,
using DFT+U\cite{anisimov_band_1991} 
and DFT+DMFT calculations.
We show that at the DFT level both Fe$_{\text{Li}}$
and  Fe$_{\text{Se}}$ exhibit a strong
tendency to magnetism, leading to a 
double-AFM ground state in which both layers are either metallic
or insulating at the same time. 
Magnetism appears extremely fragile, with
many almost-degenerate configurations competing with 
the ground state one. This indicates a strong tendency to a
{\em paramagnetic} behavior, which we describe using
dynamical mean field theory (DMFT).
We find that in this regime the behavior of the two Fe atoms
is qualitatively very different, due to the different 
charge state: 
Fe$_{\text{Li}}$ is an incipient Mott insulator,
 while Fe$_{\text{Se}}$ is fully into the Hund's metal regime.
As a result, Fe$_{\text{Li}}$ states are almost entirely removed from 
the Fermi level, while Fe$_{\text{Se}}$ bands retain a strongly coherent
character and form a typical FeSC Fermi surface.

This paper is organized as follows: In Sec.~\ref{sec:method} we report the computational details of our calculations. 
Section~\ref{sec:electronic_structure} contains 
the results of our DFT and DFT+U calculations
in the non-magnetic and magnetic regime.
In Sec.~\ref{sec:dmft} we present calculations including
correlations within DFT+DMFT. We conclude and summarize our findings in
Sec.~\ref{sec:conclusions}.

\section{Computational Details}
\label{sec:method}

\begin{figure}
  \begin{center}
     \includegraphics[width=\linewidth]{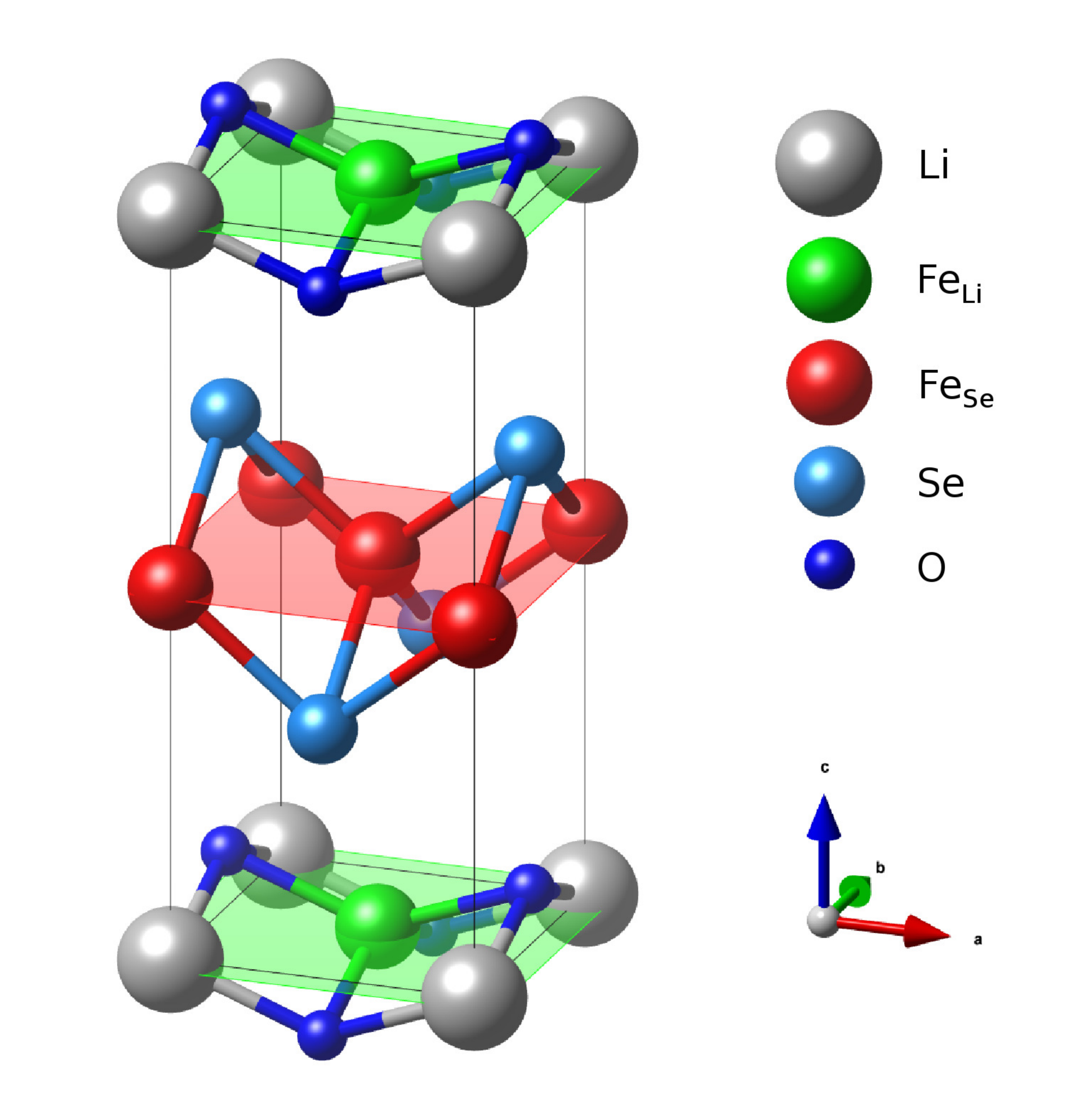}
    \caption{(Color online) Nonmagnetic unit cell of LiFeO$_2$Fe$_2$Se$_2$. The two different iron atoms are shown with two different colors, Fe$_{\text{Li}}$ in green and Fe$_{\text{Se}}$ in red. The Fe$_{\text{Li}}$ and Fe$_{\text{Se}}$ planes
are shaded in green and red, respectively.}
    \label{fig:pm_structure}
  \end{center}
\end{figure}

According to Lu~\etal~\cite{lu_superconductivity_2014} LiFeO$_2$Fe$_2$Se$_2$ crystallizes in a simple tetragonal unit cell with $a=b=3.7926$\,\r{A}, $c=9.2845$\,\r{A}, and $\alpha=\beta=\gamma=90^\circ$, which belongs to the P4/nmm space group and contains one \mbox{formula unit (\textit{f.u.}).} Each unit cell 
comprises two different types of layers: the FeSe layer common to all FeSCs, and a LiFeO$_2$ layer,
in which Li and Fe are randomly distributed on a square lattice and O atoms tetrahedrally coordinated to them.
Fe atoms in the LiFeO$_2$ - Fe$_{\text{Li}}$ - and FeSe layers - Fe$_{\text{Se}}$ - are inequivalent. They are shown in green
and red in Fig.~\ref{fig:pm_structure}, respectively. 
For all our calculations we assumed a regular, alternating in-plane arrangement of the Li and Fe$_{\text{Li}}$ atoms, so that
Li and Fe$_{\text{Li}}$ sit on top of Fe$_{\text{Se}}$ atoms. In this configuration, 
Fe$_{\text{Se}}$ occupies $4e$ and Se $8j$
Wyckoff positions with $z=0.6645$;
Li and Fe$_{\text{Li}}$ occupy $4d$ positions,
 O $8j$ positions with $z=0.0764$.
We want to note here that the Fe concentration in the LiFeO$_2$ layer is only half that of the one in the FeSe layer; thus the average nearest-neighbor Fe-Fe distance in the LiFeO$_2$ layer is a factor of $\sqrt{2}$ larger.
The FeSe tetrahedra are strongly elongated; in fact, the distance of the Se atoms from the Fe planes is $h_{Se} \approx 1.53$\,\r{A}, 
much larger than $h_{\text{Se}} \approx 1.45$\,\r{A} in bulk FeSe at zero pressure~\cite{margadonna_pressure_2009}.

For all our electronic structure calculations we have employed the full-potential linearized augmented plane-wave package WIEN2k~\cite{blaha_wien2k_2001} using a GGA-PBE exchange-correlation functional.\cite{perdew_generalized_1996}$^,$\footnote{RKmax was set to $7.0$ and the following MT radii were chosen: $\text{RMT(Li)} = 1.90 a_0$, RMT(Fe$_{\text{Li}}) = 2.02 a_0$, RMT(Fe$_{\text{Se}}) = 2.02 a_0$, $\text{RMT(Se)} = 2.15 a_0$ and $\text{RMT(O)} = 1.79 a_0$. For the reciprocal k-space integration we took \mbox{$720$ $k$-points} in the irreducible wedge.}
For the DFT+DMFT calculations we use the charge self-consistent
implementation of the TRIQS
toolkit.\cite{ferrero_triqs_????,aichhorn_dynamical_2009,aichhorn_importance_2011} As
impurity solver we employ continuous-time quantum
Monte-Carlo.\cite{werner_continuous-time_2006,werner_hybridization_2006,boehnke_orthogonal_2011}

\section{Electronic Structure}
\label{sec:electronic_structure}

\begin{figure*}
  \begin{center}
    \includegraphics[width=0.9\linewidth]{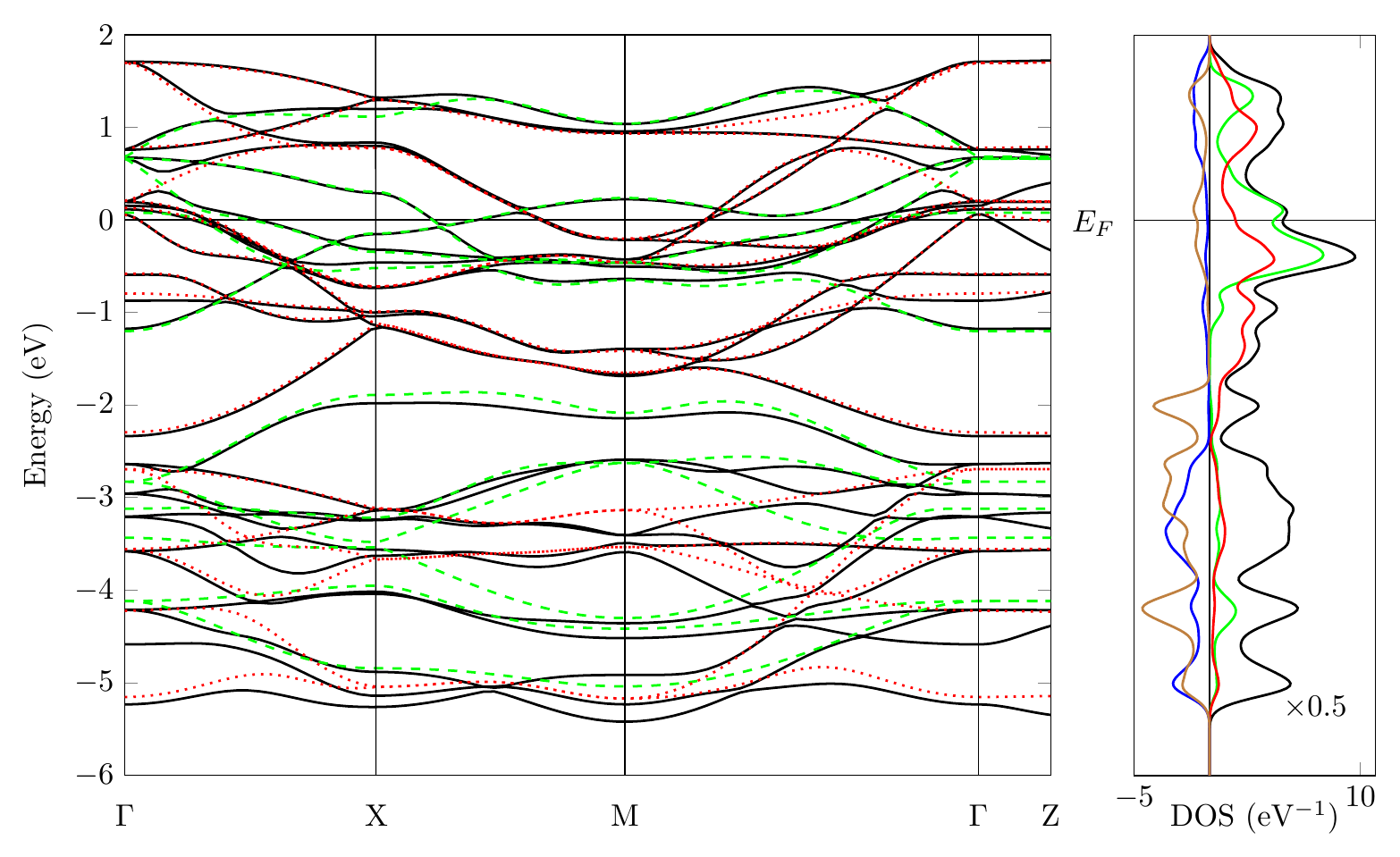}
    \caption{(Color online) Left: DFT band structure of nonmagnetic LiFeO$_2$Fe$_2$Se$_2$ (solid black lines), isolated LiFeO$_2$ (dashed green lines), and isolated FeSe (red dotted lines). Right: (p)DOS of nonmagnetic LiFeO$_2$Fe$_2$Se$_2$. 
The pDOS of Fe$_{\text{Li}}$ is shown in green and the pDOS of Fe$_{\text{Se}}$ in red. The contributions from Se and O are plotted in blue and brown, respectively, on the negative axis. 
The units are st/eV (two spins) per atom for the pDOS; the total DOS
is in st/eV f.u. (two spins) and has been rescaled by a factor of 
$2$ to improve the readability of the figure.}
    \label{fig:lda_dos}
  \end{center}
\end{figure*}

\begin{figure}
  \begin{center}
     \includegraphics[width=5.5\linewidth]{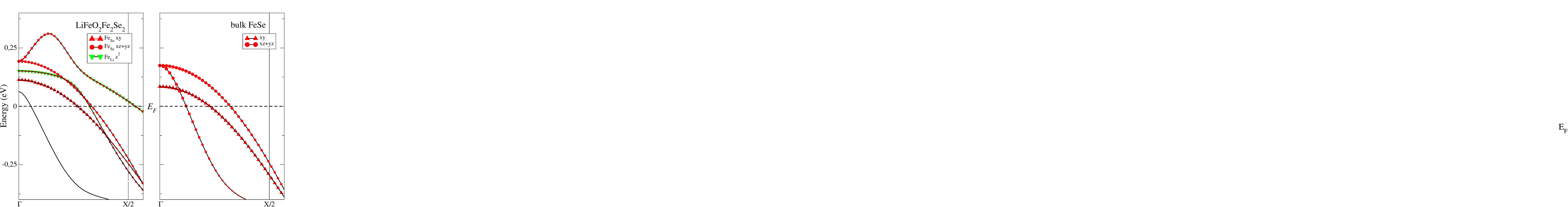}
    \caption{(Color online) Left: Zoom of Fig.~\ref{fig:lda_dos} with the bands of LiFeO$_2$Fe$_2$Se$_2$ in solid black. The dominant character of the LiFeO$_2$Fe$_2$Se$_2$ bands is shown by the symbols. The band without special character markers is a mixture of all characters. Right: Band structure of bulk FeSe calculated for the experimental crystal structure at ambient pressure from Ref.~\cite{margadonna_pressure_2009}.}
    \label{fig:pm_bandstructure_zoom}
  \end{center}
\end{figure}
Figure~\ref{fig:lda_dos} shows the non-magnetic DFT bandstructure (left panel) and the (partial) density of states (pDOS) of LiFeO$_2$Fe$_2$Se$_2$ (right panel), in an energy range of $[-6,2]$\,eV around the Fermi level, which is chosen at zero energy. The bandwidth of Fe$_{\text{Li}}$ is about $2$\,eV, which is due to the larger average Fe-Fe distance in this layer, and that of Fe$_{\text{Se}}$ $3.5$\,eV. States at the Fermi level have mostly Fe$_{\text{Li}}$ and Fe$_{\text{Se}}$ partial character, while ligand (O, Se) bands lie lower in energy. The peak at approximately $-2$\,eV has mainly O character and it is clearly seen that around $-4.1$\,eV there is a strong hybridization between Fe$_{\text{Li}}$ and O, while Fe$_{\text{Se}}$ and Se show hybridization between $-3.9$\,eV and $-2.5$\,eV.
The DOS at the Fermi energy is \mbox{$N(E_F) = 7.7$\,eV$^{-1}$} \textit{f.u.} per spin,
i.e., above the Stoner criterion.

The left panel of Fig.~\ref{fig:lda_dos} shows the corresponding electronic structure along high-symmetry
lines in the Brillouin zone (BZ). Our results agree with those of Ref.~\onlinecite{liu_coexistence_2014}.
In addition to the bands of the full compound (solid black lines), we also show the bands of the {\em isolated} LiFeO$_2$ (dashed green lines) 
and FeSe layers (dotted red lines) in the original unit cell. 
In order to align the bands, we had to
shift the bands of isolated LiFeO$_2$  down by $-0.1$\,eV,
which corresponds to a charge transfer of $0.35$\,e$^-$ from FeSe to LiFeO$_2$ layers. 
Except for an increased dispersion of the Fe$_{\text{Se}}$ bands along the $\Gamma$-Z direction, the low-energy band structure of LiFeO$_2$Fe$_2$Se$_2$ coincides almost exactly with that of the isolated layers.

Due to the strong elongation of the Fe-Se tetrahedra and to hybridization
with the LiFeO$_2$ states, the fermiology  of LiFeO$_2$Fe$_2$Se$_2$ is quite different compared to typical iron-chalcogenide SCs.
In Fig.~\ref{fig:pm_bandstructure_zoom}, we show the low-energy bandstructure of LiFeO$_2$Fe$_2$Se$_2$ and bulk FeSe decorated with the dominant orbital characters along a short section of the $\Gamma$-X path. This permits us to 
highlight and understand the difference in the shape of the hole pockets.

The Fe$_{\text{Se}}$ 3$d_{xy}$ band close to the Fermi energy (red
triangles in Fig.~\ref{fig:pm_bandstructure_zoom}) has the same
dispersion in LiFeO$_2$Fe$_2$Se$_2$ and bulk FeSe. 
On the other hand, the remaining hole bands have very different dispersions in the two compounds. In particular, one of the doubly degenerate $d_{xz/yz}$ bands which form the hole pockets in most FeSCs is pushed up in LiFeO$_2$Fe$_2$Se$_2$ due to the large $h_{Se}$,~\cite{kuroki_unconventional_2008,andersen_multi-orbital_2011} and is further modified by hybridizations with Fe$_{\text{Li}}$ $d_{z^2}$. These hybridizations are so strong that when this band crosses $E_F$, it has mostly Fe$_{\text{Li}}$ character.
As one $d_{xz/yz}$ band is removed from the Fermi surface, another band appears at $E_F$. This new band is mainly a mixture of Fe$_{\text{Se}}$ $d_{xz/yz}$ and Fe$_{\text{Li}}$ $d_{z^2}$.

\begin{figure}
  \begin{center}
     \includegraphics[width=1.0\linewidth]{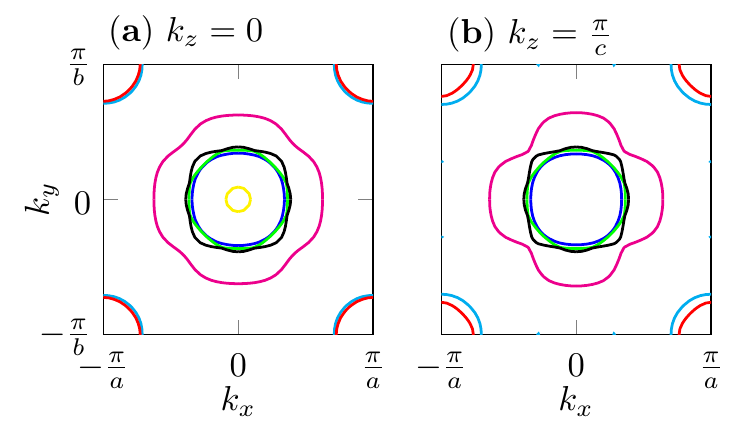}
    \caption{(Color online) Top view of the Fermi surface at $k_z=0$ (left) and $k_z=\pi/c$ (right) of non-magnetic LiFeO$_2$Fe$_2$Se$_2$. Different colours indicate different bands.} 
    \label{fig:pm_fs}
  \end{center}
\end{figure}

Figure~\ref{fig:pm_fs} shows the Fermi surface of LiFeO$_2$Fe$_2$Se$_2$ in the \mbox{$k_z=0$} and the \mbox{$k_z=\pi/c$} planes. 
In this figure, different colors indicate different bands and are
not related to orbital character.
The smallest hole pocket has a three-dimensional cigar shape
and is located around the $\Gamma$ point [yellow line in Fig.~\ref{fig:pm_fs}(a) and not present in Fig.~\ref{fig:pm_fs}(b)].
The other hole pockets are shown in blue (Fe$_{\text{Se}}$ $d_{xy}$),
green (Fe$_{\text{Li}}$/Fe$_{\text{Se}}$), and black (Fe$_{\text{Se}}$ 3$d_{xz+yz}$).
The electron pockets at the M points, shown in cyan and red,
have mostly $d_{xz/yz/xy}$ character and
are much less affected by hybridization and changes in selenium height $h_{Se}$. 
In addition to the FeSe pockets,
the LiFeO$_2$ layer provides an additional hole pocket in the middle of the Brillouin zone (magenta lines in Fig.~\ref{fig:pm_fs}),
which has a considerable three-dimensional character.

\begin{figure}
  \begin{center}
\includegraphics[width=1.0\linewidth]{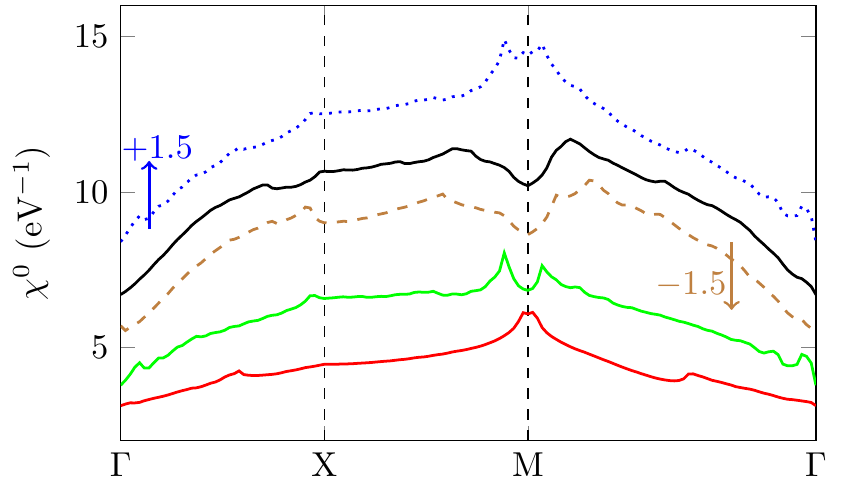}
\end{center}
\caption{(Color online) Static bare susceptibility $\chi^0$ for isolated FeSe (red), isolated LiFeO$_2$ (green), and the full LiFeO$_2$Fe$_2$Se$_2$ compound (black). The dotted blue line results when summing $\chi^0$ for the two isolated layers. The dashed brown line depicts the summed $\chi^0$ of the isolated layers, where we performed a rigid-band shift in order to account for the charge transfer. All susceptibilities are given per spin and for all Fe atoms in the respective unit cells. The blue and brown curves have been shifted up (down) by 1.5 eV$^{-1}$
to improve readability.
For details on the susceptibility calculation see Ref.~\onlinecite{heil_accurate_2014}.}
\label{fig:rechi0}
\end{figure}

The charge transfer between the layers and the presence of the
additional LiFeO$_2$-derived band are quite visible also 
in the susceptibility $\chi^0$, plotted in Fig.~\ref{fig:rechi0}.
Details of the calculations are given in Ref.~\onlinecite{heil_accurate_2014}.
The susceptibility of the full compound, shown as a black
solid line, grows towards the border of the
Brillouin zone (X-M line), and shows a dip around the M point,
in contrast to most FeSCs, which show a clear maximum at M. 
Indeed, the isolated FeSe layer (red curve) shows a well-defined peak
 at this point. The LiFeO$_2$ layer (green solid line) has an even 
larger susceptibility, with a dip
around the M point. In order to explain the full susceptibility
  it is not sufficient to sum the contributions from the isolated
  layers, which still shows a maximum
  around the M point (dotted blue curve). In order for the sum of the two layers to
  reproduce the susceptibility, the Fermi levels have to be adjusted  
as done in Fig.~\ref{fig:lda_dos}. This results in the dashed brown curve of
Fig.~\ref{fig:rechi0} and highlights the occurrence of charge transfer in this material.

We now discuss results of spin-polarized DFT calculations for the magnetically ordered states, which are shown in Tables~\ref{tab:isolated_energies} and \ref{tab:life02fe2se2_energies}. We considered the most important magnetic configurations for the isolated layers and the full compound. In Table~\ref{tab:isolated_energies} we show the results for the magnetic moments ($m_\mathrm{Fe}$) and the stabilization energy per Fe atom ($\Delta E$) for ferromagnetic (fm), checkerboard (cb), single stripe (ss), and double stripe (ds) configurations for the isolated layers.
\begin{table}
  \begin{center}
    \caption{Energies (with respect to the non-magnetic configuration) and magnetic moments of isolated LiFeO$_2$ $m(\mathrm{Fe_{Li}})$ and isolated FeSe $m(\mathrm{Fe_{Se}})$ for ferromagnetic (fm), checker-board (cb), single stripe (ss), and double stripe (ds) magnetic configurations.}
    \begin{tabular}{ l   c   c   c  }
    \hline \hline
     isolated LiFeO$_2$    & $\Delta E$/Fe (meV)    & $m(\mathrm{Fe_{Li}})$ ($\mu_B$) \\ \hline
     fm                    & -1358.12               & 3.79  \\
     cb                    & -1693.55               & 3.57  \\
     ss                    & -1472.59               & 3.55  \\
%
     isolated FeSe         & $\Delta E$/Fe (meV)    & $m(\mathrm{Fe_{Se}})$ ($\mu_B$) \\ 
     fm                    & -274.56                & 2.38 \\
     cb                    & -511.29                & 2.29 \\
     ss                    & -564.28                & 2.51 \\
     ds                    & -423.70                & 2.53 \\ \hline \hline
    \end{tabular}
    \label{tab:isolated_energies}
  \end{center}
\end{table}
In isolated LiFeO$_2$ we find that the cb order is the one
with the lowest energy, with an energy difference of $221$\,meV from
the ss and $335$\,meV from the fm configuration.
In the isolated FeSe compound the ss magnetic order is the
most favourable one, separated from the cb order by $53$\,meV
and from the ds order by $141$\,meV.
Fe$_{\text{Se}}$ and Fe$_{\text{Li}}$ have different magnetic moments;
the filling of the atoms implies a saturation moment of 4 and 5~$\mu_B$, respectively. In fact, we find that $m$ of Fe$_{\text{Se}}$ is 
$2.4$\,$\mu_B$~\footnote{This is in line with DFT calculations for
other FeSCs in $d^6$ configuration~\cite{subedi_density_2008,yin_kinetic_2011}. This
larger $m(\mathrm{Fe_{Se}})$ is a consequence of the large Se height.},
while the magnetic moment of Fe$_{\text{Li}}$ in LiFeO$_2$ is $\sim
3.6$\,$\mu_B$. 
These values are almost independent of the magnetic ordering pattern,
both in the isolated layers and in the full compound.

\begin{figure}
  \begin{center}
     \includegraphics[width=0.95\linewidth]{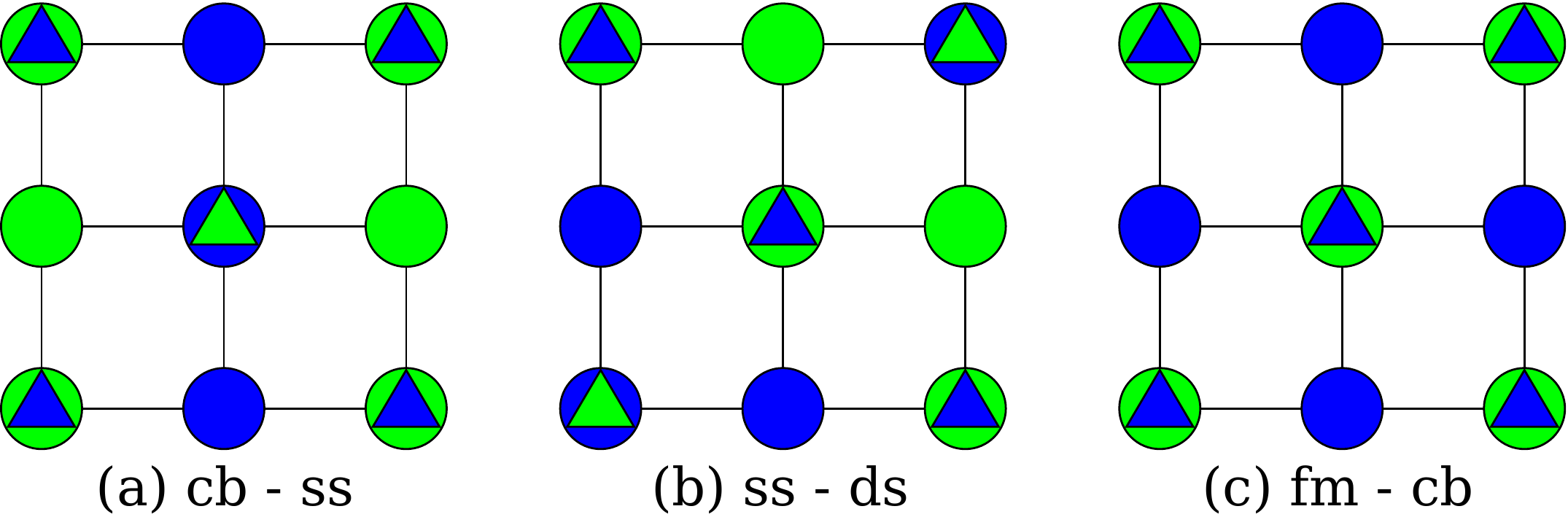}
    \caption{(Color online) Top view of the Fe lattices in LiFeO$_2$Fe$_2$Se$_2$ for the three most preferable magnetic configurations in terms of energy. The circles represent the Fe$_{\text{Se}}$ and the triangles Fe$_{\text{Li}}$. Blue signifies an up-spin on that particular Fe atom and green a down-spin.} 
    \label{fig:magnetic_moments_lifeo2fe2se2}
  \end{center}
\end{figure}

\begin{table}
  \begin{center}
    \caption{Energies (with respect to the nonmagnetic configuration) and magnetic moments of LiFeO$_2$Fe$_2$Se$_2$ in different magnetic configurations (compare with Fig.~\ref{fig:magnetic_moments_lifeo2fe2se2}).}
    \begin{tabular}{ l   c   c   c  }
    \hline \hline
     LiFeO$_2$Fe$_2$Se$_2$ & $\Delta E$/Fe (meV)    & $m(\mathrm{Fe_{Li}})$ ($\mu_B$) & $m(\mathrm{Fe_{Se}})$ ($\mu_B$) \\ \hline
     fm - fm               & -247.90                & 3.71 & 1.77 \\
     cb - ss               & -738.25                & 3.55 & 2.40 \\
     ss - ds               & -655.25                & 3.60 & 2.38 \\
     fm - cb               & -575.29                & 3.81 & 2.21 \\ \hline \hline
    \end{tabular}
    \label{tab:life02fe2se2_energies}
  \end{center}
\end{table}

Table~\ref{tab:life02fe2se2_energies} reports the most stable 
configurations for the full compound;
the corresponding patterns are also shown in Fig.~\ref{fig:magnetic_moments_lifeo2fe2se2}.
 Note that, since
the two Fe sublattices are rotated by $45^\circ$ with
respect to each other, and the reciprocal unit cell of the Fe$_{\text{Li}}$
sublattice is smaller, we have $\mathbf{Q}_{\text{cb,ss}} =
(\pi,\pi,0)$, $\mathbf{Q}_{\text{ss,ds}} = (\pi,0,0)$, and
$\mathbf{Q}_{\text{fm,cb}} = (2n\pi,0,0)$; i.e.,
the most stable configurations are those in which the
ordering vectors of the two sublattices are commensurate.
In particular, we find that 
the configuration with the
lowest energy is the one where the Fe$_{\text{Li}}$ have \textit{cb} order and
the Fe$_{\text{Se}}$ are aligned in an \textit{ss} way [see
Fig.~\ref{fig:magnetic_moments_lifeo2fe2se2}(a)]. This is in agreement
with what was reported by Liu~\etal~\cite{liu_coexistence_2014}.
The magnetic coupling {\em between} the FeSe and LiFeO$_2$ planes
is extremely weak; indeed, we find that an AFM alignment of the
spins along the $z$ direction is slightly favorable, but the
energy difference from the FM case is $\lesssim 3$ meV. 
In addition to the configurations reported in the table, we also found
many metastable ones, indicating a very fragile
nature of magnetism and a strong tendency to magnetic fluctuations
in this compound.

\begin{figure}
  \begin{center}
    \includegraphics[width=1.0\linewidth]{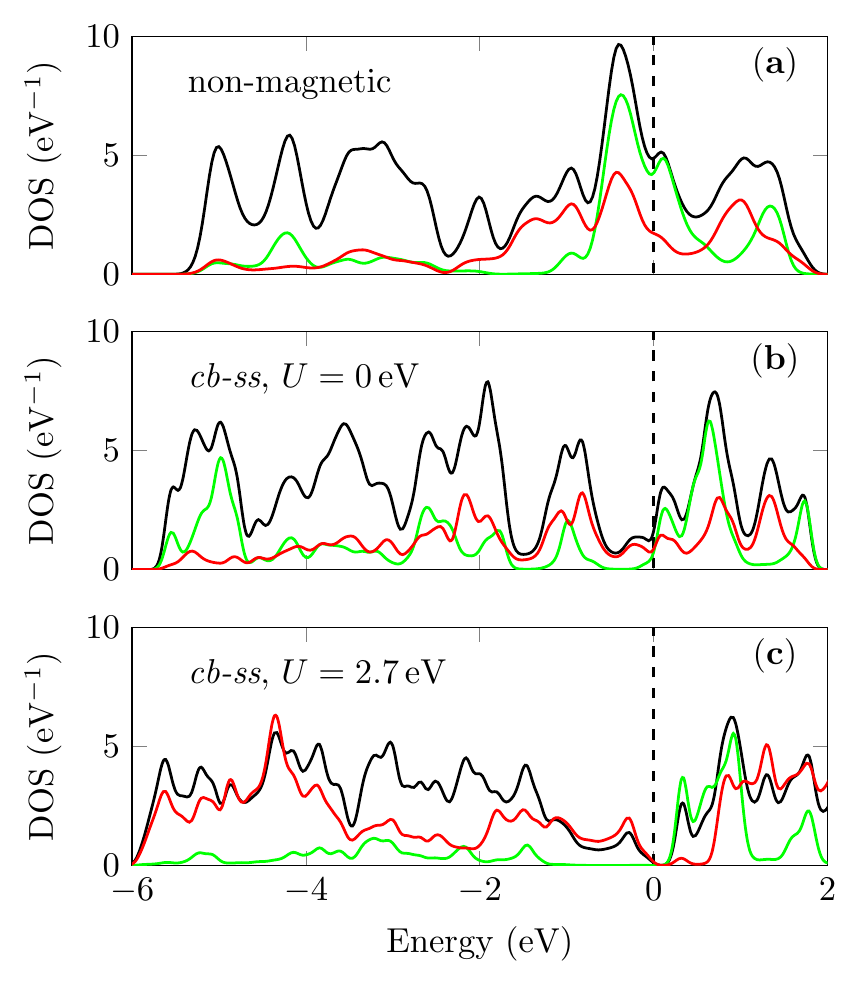}
    \caption{(Color online) 
      \mbox{$\mathbf{(a)}$: (p)DOS} of non-magnetic LiFeO$_2$Fe$_2$Se$_2$. The
      total DOS is
      shown in black  and the pDOS of Fe$_{\text{Li}}$ in green
      and of Fe$_{\text{Se}}$ in red (same as Fig.~\ref{fig:lda_dos}).
      \mbox{$\mathbf{(b)}$: (p)DOS} of cb-ss
      LiFeO$_2$Fe$_2$Se$_2$. The total DOS is
      shown in black and the pDOS of Fe$_{\text{Li}}$ in green
      and of Fe$_{\text{Se}}$ in red, where we summed over majority and minority spins.
      \mbox{$\mathbf{(c)}$: (p)DOS} of cb-ss
      LiFeO$_2$Fe$_2$Se$_2$ and U$=2.7$\,eV. Otherwise the same as $\mathbf{(b)}$. Units are the same as in Fig.~\ref{fig:lda_dos}. 
      } 
    \label{fig:cbss_lsda_dos}
  \end{center}
\end{figure}

In all the cases we considered, both layers in LiFeO$_2$Fe$_2$Se$_2$
are metallic in spin-polarized DFT. \mbox{Panel (b)} of Fig.~\ref{fig:cbss_lsda_dos}
depicts the total and partial DOS of LiFeO$_2$Fe$_2$Se$_2$
in the \textit{cb-ss} configuration, compared with the non-magnetic \mbox{one (a)}.
The figure clearly shows a considerable depletion
of the spectral weight at the Fermi level in both layers,
but both Fe$_{\text{Se}}$ (red) and Fe$_{\text{Li}}$ (green) contribute states at the Fermi level.
Liu~\etal~\cite{liu_coexistence_2014} 
have argued that a small Coulomb interaction $U$
would be sufficient to open a gap at the Fermi level for
Fe$_{\text{Li}}$, but not for Fe$_{\text{Se}}$.
To check this hypothesis, we performed DFT+U calculations 
for several values of $U$, and found that a gap
opens {\em simultaneously} in the two layers, 
for values of $U > 2$ eV. A calculation for $U = 2.7$ eV is reported in panel (c)
of Fig.~\ref{fig:cbss_lsda_dos}, and shows fully developed gaps for
both layers.

\begin{figure}
  \begin{center}
\includegraphics[width=1.0\linewidth]{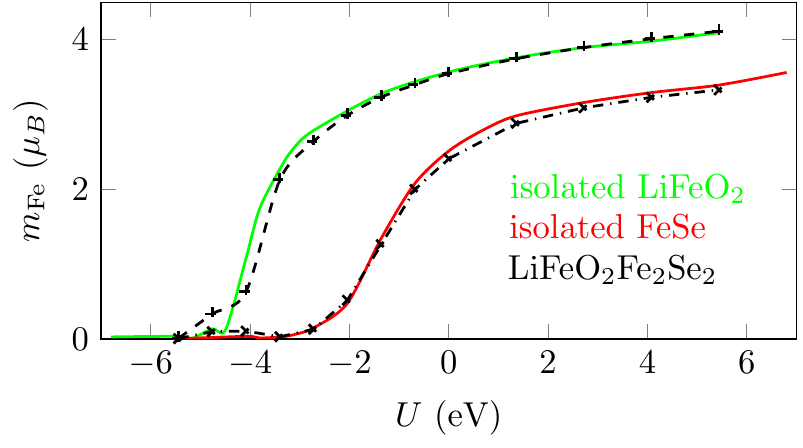}
\end{center}
\caption{(Color online) Magnetic moment of Fe in ss isolated FeSe (red), cb isolated LiFeO$_2$ (green) and of Fe$_{\text{Li}}$ (dotted black, $+$) and Fe$_{\text{Se}}$ (dash-dotted black, $\times$) in cb-ss LiFeO$_2$Fe$_2$Se$_2$.}
\label{fig:afm_mofU}
\end{figure}

In Fig.~\ref{fig:afm_mofU} we plot the values of the
magnetic moments  obtained in DFT+U calculations.
The results for the full compound are shown as black symbols; they
almost perfectly match those obtained for the isolated layers,
shown in red and green, respectively, confirming the small
coupling between the two layers. 
Note that $U$ in this figure ranges from $-7$ to $7$\,eV;
positive values of $U$ have a clear physical meaning, while 
``negative $U$'' DFT+U calculations have
been introduced in the early days of FeSCs as a phenomenological way
to simulate the reduction of the magnetic moment due to spin
fluctuations~\cite{ferber_analysis_2010}.
We will use them in this context only to visually characterize how
robust the magnetism is.

Several observations are in place at this point: 
the saturation values are clearly different for LiFeO$_2$ and FeSe, 
but they are not reached for $U = 7$\,eV, indicating that charge
fluctuations are important in both compounds.
Magnetism is much more robust in LiFeO$_2$, since 
it takes much larger negative values of $U$ to suppress it; however,
the suppression is much faster, once the critical $U$ is
approached. It is possible that in the full compound the effective
value of the Coulomb interaction is different in the two layers, due to the different nature
of the ligands and different Fe-Fe distance. 
However, in order to
recover (within DFT+U) a solution with no long-range magnetic order in the FeSe layer and
an insulating LiFeO$_2$ layer it would be necessary to assume negative
$U$ values for Fe$_{\text{Se}}$ and positive values for
Fe$_{\text{Li}}$.
This indicates that DFT+U is not able to describe this system consistently. 
An alternative description, which takes into account the dynamical nature of correlations,
is given in the following section.

\section{Correlated electronic structure}
\label{sec:dmft}

\begin{figure*}
  \begin{center}
     \includegraphics[width=0.9\linewidth]{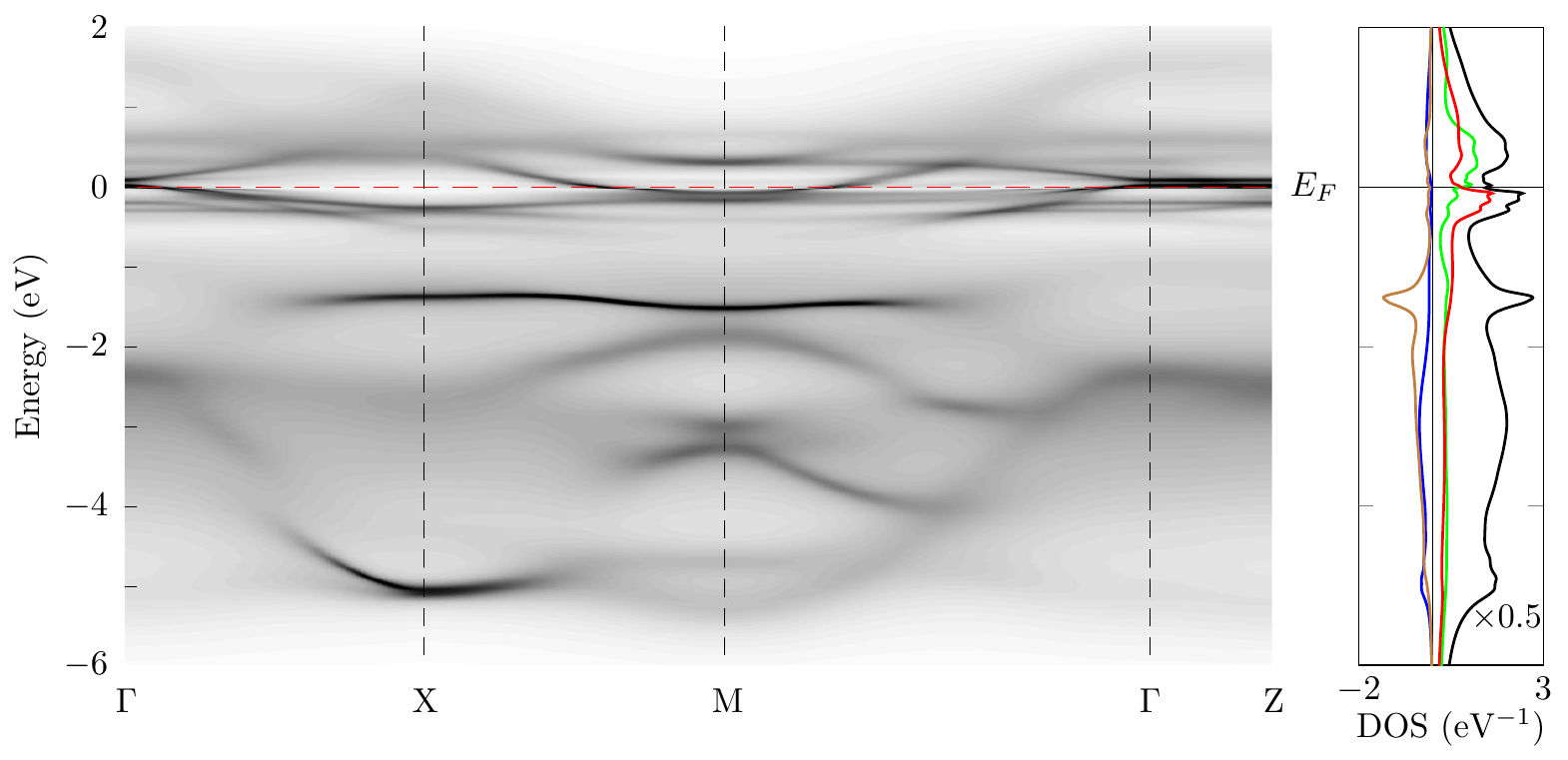}
    \caption{(Color online) Left: DFT+DMFT spectral function of paramagnetic LiFeO$_2$Fe$_2$Se$_2$ for \mbox{$U=4$\,eV} and \mbox{$J=0.9$\,eV}. Right: DFT+DMFT DOS of paramagnetic LiFeO$_2$Fe$_2$Se$_2$ for the same $U$ and $J$ parameters.  The pDOS of Fe$_{\text{Li}}$ is shown in green and the pDOS of Fe$_{\text{Se}}$ in red. The contributions from Se and O are plotted on the negative axis and are displayed in blue and brown, respectively.
For the definition of units, see Fig.~\ref{fig:lda_dos}.} 
    \label{fig:dmft_dos_bands}
  \end{center}
\end{figure*}

\begin{figure}
  \begin{center}
     \includegraphics[width=1.0\linewidth]{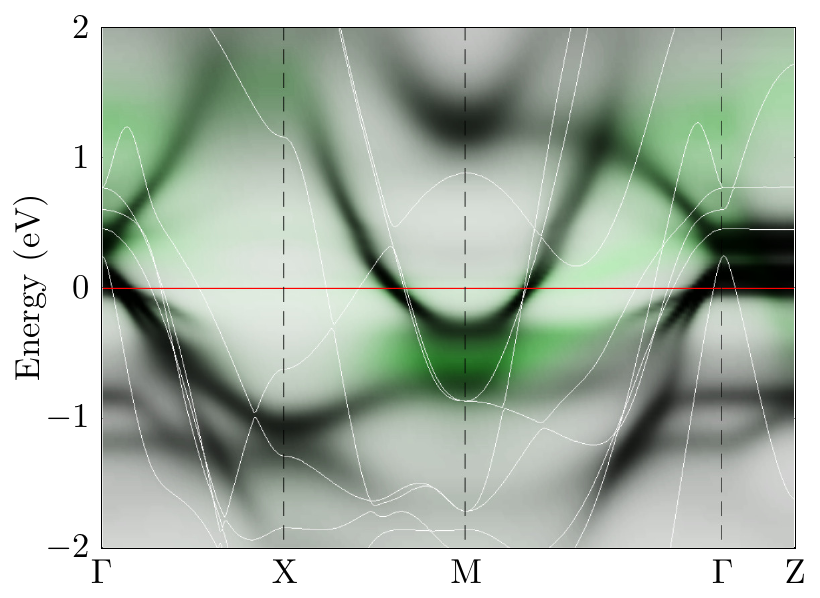}
    \caption{(Color online) DFT+DMFT spectral function of paramagnetic LiFeO$_2$Fe$_2$Se$_2$ for $U=4$\,eV and $J=0.9$\,eV. The Fe$_{\text{Li}}$ 3$d_{z^2}$ contributions are drawn in green and the DFT band structure is overlaid in white.} 
    \label{fig:akw_pm_J09}
  \end{center}
\end{figure}

\begin{figure}
  \begin{center}
      \includegraphics[width=0.95\linewidth]{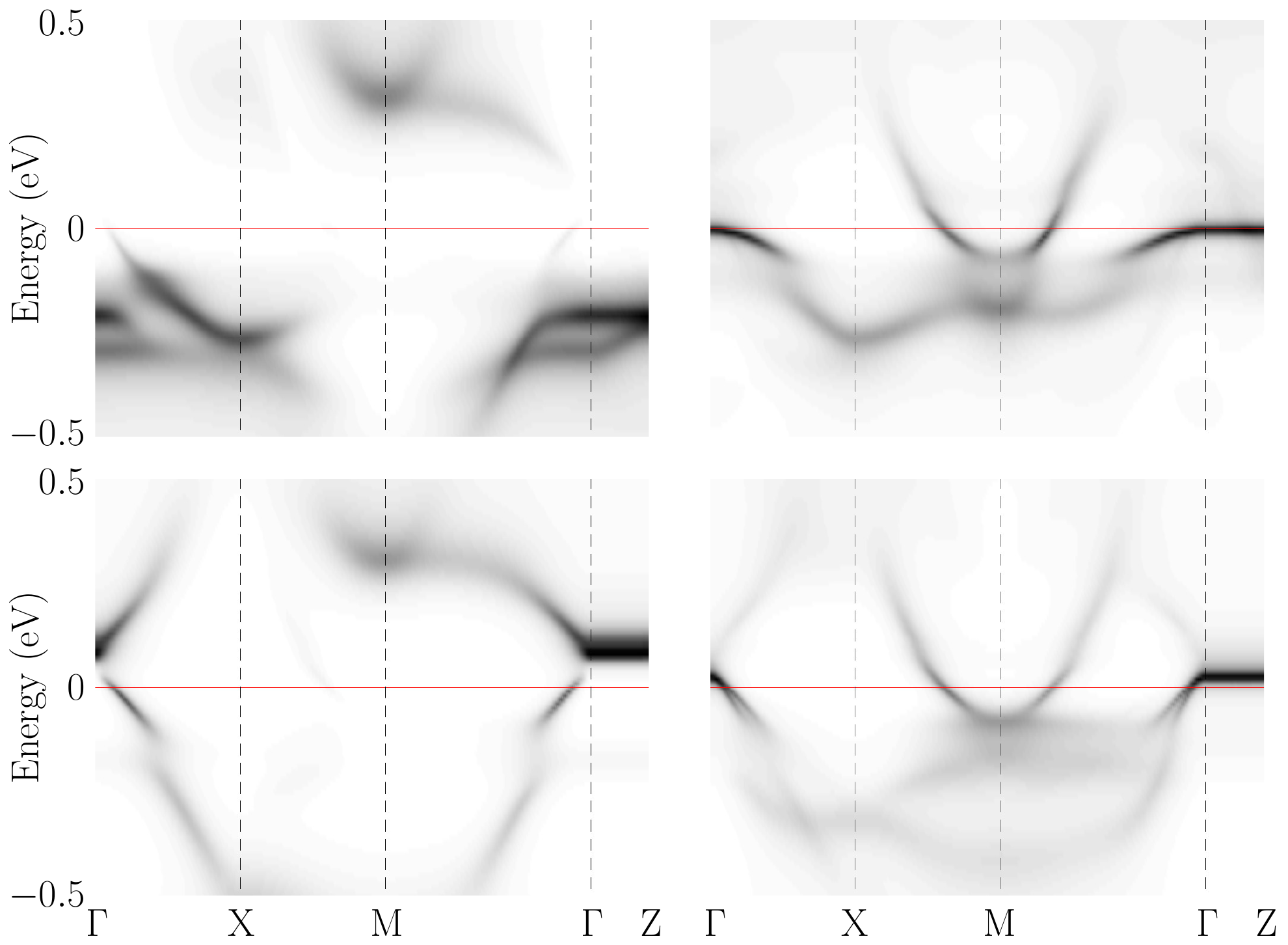}
    \caption{(Color online) Contributions of Fe$_{\text{Se}}$ to the spectral function of paramagnetic LiFeO$_2$Fe$_2$Se$_2$ for \mbox{$U=4$\,eV} and \mbox{$J=0.9$\,eV} separated into orbital contributions. Top left: Fe$_{\text{Se}}$ 3$d_{z^2}$ orbital, top right: Fe$_{\text{Se}}$ 3$d_{x^2+y^2}$, bottom left: Fe$_{\text{Se}}$ 3$d_{xy}$, bottom right: Fe$_{\text{Se}}$ 3$d_{xz+yz}$.} 
    \label{fig:akw_orbital_pm_J09}
  \end{center}
\end{figure}

As mentioned in the beginning, the two iron atoms have very different 
properties related to their nominal charge. For a half-filled atomic shell, as in Fe$_{\text{Li}}$, the Hund's rule coupling enhances correlation effects
resulting in an insulating behavior, whereas off half filling, as in Fe$_{\text{Se}}$, Hund's metallicity (small coherence scale but no Mott transition) shows up~\cite{de_medici_janus-faced_2011,georges_strong_2013,yin_kinetic_2011}.
Since in LiFeO$_2$Fe$_2$Se$_2$ we have iron atoms with the two valences {\em in one single compound}, it is interesting to study their response to 
correlations in DMFT, and to see whether the general arguments given above hold here.

For the band structure calculations, we again use the WIEN2k
code package. For the treatment of correlations we apply
the continuous-time quantum Monte Carlo technique in the hybridization
expansion formulation~\cite{werner_hybridization_2006,werner_continuous-time_2006}, as implemented in the TRIQS package~\cite{ferrero_triqs_????,boehnke_orthogonal_2011}. We use full charge self-consistency~\cite{aichhorn_importance_2011}, as well as spin-flip and pair-hopping terms in the local Hamiltonian~\cite{parragh_conserved_2012}. Wannier
functions are 
constructed from Fe $d$, Se $p$, and O $p$ states within an energy window of
$[-6,2.5]$\,eV.
Consistent with previous work on
bulk FeSe~\cite{aichhorn_theoretical_2010} we choose $U = 4$\,eV and $J =
0.9$\,eV as interaction values, for both Fe$_{\text{Li}}$ and Fe$_{\text{Se}}$ atoms.

The spectral function and the (p)DOS of paramagnetic
LiFeO$_2$Fe$_2$Se$_2$ can be seen in
Fig.~\ref{fig:dmft_dos_bands}. The left panel clearly shows a shrinking
of the bandwidth of the bands around the Fermi level and that
excitations become incoherent already at rather low binding energies
of around $0.4$\,eV. From $-5.5$\,eV to $-2$\,eV the bands have
mainly O and Se character, with only very little Fe contributions. The
sharp peak around $-1.4$\,eV has solely O character.

In order to disentangle the effect of correlations on Fe$_{\text{Li}}$ and Fe$_{\text{Se}}$
atoms we show a close-up of the spectral function around the Fermi
level in Fig.~\ref{fig:akw_pm_J09}.
The contributions from Fe$_{\text{Li}}$ to the spectral function are drawn in green,
while the Fe$_{\text{Se}}$ bands are shown in black. It is immediately clear that
the sharp features in the band structure,
corresponding to well-defined quasiparticles,
 stem solely from the Fe$_{\text{Se}}$
atom, which has a 3$d^6$ configuration and is hence in the Hund's
metallic regime. In particular, when comparing with the DFT band
structure (Fig.~\ref{fig:lda_dos}, redrawn in
Fig.~\ref{fig:akw_pm_J09} as white lines) one can easily see that
correlations remove all coherent contributions of the Fe$_{\text{Li}}$ atom to
the low-energy band structure.
The only contributions from Fe$_{\text{Li}}$ that survive
in the vicinity of $E_F$ are at the $\Gamma$ and M points and are of
3$d_{z^2}$ character. They have, however, very incoherent character and are thus heavily smeared out. 
The 
remaining bands from Fe$_{\text{Se}}$ restore a very familiar picture of the
Fermi surface topology: three hole pockets at the $\Gamma$ point and
two electron pockets near the M points.

The Fe$_{\text{Se}}$ orbital character of the bands is shown in
Fig.~\ref{fig:akw_orbital_pm_J09}. 
While the Fe$_{\text{Se}}$ 3$d_{z^2}$ has distinct
features exclusively below the Fermi energy and the Fe$_{\text{Se}}$
3$d_{x^2+y^2}$ contributes only slightly to the middle hole pocket,
the main parts of the Fermi surface come from Fe$_{\text{Se}}$ 3$d_{xy}$ and
3$d_{xz/yz}$. The outer hole pocket has almost only Fe$_{\text{Se}}$ 3$d_{xy}$
character, while the two inner hole pockets have 3$d_{xz/yz}$
character. The electron pocket at the M point is a combination of
those two, with 3$d_{xz/yz}$ being the stronger one. 
The mass enhancements of these orbitals are between $\sim 2$ for the e$_g$ orbitals, 2.8 $d_{xz/yz}$, and 3.2 for $d_{xy}$.
Again, let us
stress that there is no coherent contribution from the Fe$_{\text{Li}}$ atom to
the bands forming the Fermi surface. 

It is interesting to compare our band structure with that of bulk FeSe. 
In Ref.~\onlinecite{aichhorn_theoretical_2010} this band structure is
shown and reveals a striking similarity with our
compound here. 
In agreement with bulk FeSe~\cite{aichhorn_theoretical_2010} we
have 3 hole pockets at the $\Gamma$ point, where the outermost
pocket is of dominantly $d_{xy}$ character. In both compounds we have 
orbital-dependent mass renormalization, with the $d_{xy}$ being the most heavy orbital.

However, differently from what was argued based on DFT+U
calculations~\cite{liu_coexistence_2014}, we do not find a strict Mott
insulator for the LiFeO$_2$ layer, when the interaction values are taken as $U=4.0$\,eV and $J=0.9$\,eV.
Instead, Fe$_{\text{Li}}$ is in a strongly
orbital-selective Mott regime~\cite{de_medici_orbital-selective_2009,yin_kinetic_2011,lanata_orbital_2013,de_medici_selective_2014}, with 4 out of 5 orbitals being
insulating, and only one (the $d_{z^2}$) with some finite, but very
incoherent, weight at 
zero energy.
An increase of the interaction values for Fe$_{\text{Li}}$ to $U=6$\,eV and 
$J=1.0$\,eV, however, results in the suppression of also the Fe$_{\text{Li}}$ $d_{z^2}$ states from the Fermi surface. 
We relate this to the fact that a 
completely incoherent state can be reached with not too large values 
for the interaction parameters due to the atomic configuration (half-filled) of the Fe$_{\text{Li}}$ atom.

However, as already discussed above, there is some intrinsic charge
transfer from Fe$_{\text{Se}}$ to Fe$_{\text{Li}}$ in this
compound. 
Calculating the charge of the
iron atoms from DFT, using $d$-only Wannier functions, gives 5.12
electrons for 
Fe$_{\text{Li}}$, which is slightly above the integer value for
half filling. That means that correlations have to overcome this small charge transfer 
and push the Fe$_{\text{Li}}$ closer to half filling, before a complete suppression of the Fe$_{\text{Li}}$ contributions can take place.

\section{Conclusions}
\label{sec:conclusions}
We presented calculations for the electronic and magnetic behavior of
the recently synthesized Fe-based superconductor LiFeO$_2$Fe$_2$Se$_2$
using first-principles DFT methods and DFT+DMFT calculations. 

The most favourable magnetic configuration in DFT has
checkerboard order in the LiFeO$_2$ and single stripe order in the
FeSe layer. When correlations in the framework of DFT+U are included,
the magnetic moments of Fe$_{\text{Li}}$ and Fe$_{\text{Se}}$ react very differently to the
Coulomb interaction $U$. While $m(\mathrm{Fe_{Li}})$ of the
LiFeO$_2$ layer is robust over a wide range of interactions and
breaks down at a negative $U$ of $-4$\,eV in a sharp transition,
the moment $m(\mathrm{Fe_{Se}})$ of the FeSe layer changes
rather smoothly with $U$.   
We find that the pDOS of Fe$_{\text{Li}}$ and Fe$_{\text{Se}}$ of cb-ss
LiFeO$_2$Fe$_2$Se$_2$ have states at the Fermi energy for $U<2$\,eV,
while for $U>2$\,eV this compound is fully gapped, in contrast to
previous reports~\cite{liu_coexistence_2014}.

We also observed that there are many magnetic configurations that are 
very close in energy to the ground state. This means that the
spins are able to fluctuate as the magnetic configuration of the
whole compound can change easily.

The nonmagnetic DFT electronic band structure is much richer compared to other
FeSCs, due to the presence of LiFeO$_2$-derived states at the Fermi level.
However, the picture is 
strongly modified when including correlation effects by means of DFT+DMFT. 
Almost all contributions from the
intercalated Fe$_{\text{Li}}$ in the LiFeO$_2$ are removed from the vicinity of
the Fermi energy with the exception of the Fe$_{\text{Li}}$ 3$d_{z^2}$ band. This
has, however, a much smaller and incoherent weight than the contributions from Fe$_{\text{Se}}$,
which means that the low-energy physics of LiFeO$_2$Fe$_2$Se$_2$ is
governed by the Fe$_{\text{Se}}$ of the FeSe layer. 
This low-energy electronic structure, stemming from Fe$_{\text{Se}}$, is very
similar to what was found for bulk FeSe~\cite{aichhorn_theoretical_2010}. Three hole pockets are located
in the vicinity of the $\Gamma$-point and two electron pockets near
the M-point, recovering the usual Fermi surface picture of Fe-based
superconductors.  
Our calculations show unambiguously that the topology of the Fermi
surface, even above a magnetic ordering temperature, should be very
similar to the well-known pocket structure of other iron-based
pnictides; this should be immediately verifiable in ARPES experiments.

The striking difference between the behavior of
Fe$_{\text{Li}}$- and Fe$_{\text{Se}}$-derived states, related to the
different valences of the two atoms,  
is one of the most spectacular realizations so far
of qualitatively different effects of Hund's rule coupling, depending
on the valence state of the atoms, in one single compound.

{\em Note:} Recently, we became aware of
another paper which discusses an alternative antiferromagnetic order
in  
FeSe monolayers and LiFeO$_2$Fe$_2$Se$_2$~\cite{cao_insulating_2014}. We note that the ordering pattern is
consistent with the slight maximum along the X-M line in the full
susceptibility in Fig.~\ref{fig:rechi0}.

\section*{Acknowledgements:}
The authors would like to thank G. Giovannetti for useful discussion. Calculations have been done on the dcluster of TU Graz.
L.B. acknowledges partial support through the DFG-SPP 1458 project, 
Grant No. Bo-3536/2. M.A. acknowledges support from the Austrian Science
Fund (FWF) through the SFB VICOM, subproject F04103.

%

\end{document}